# Accurate and efficient band gap predictions of metal halide perovskites using the DFT-1/2 method: GW accuracy with DFT expense

S. X. Tao[1*], X. Cao[1], P. A. Bobbert[1]

[1] Center for Computational Energy Research, Department of Applied Physics, Eindhoven University of Technology, the Netherlands

*Email: s.x.tao@tue.nl

**Abstract:** The outstanding optoelectronics and photovoltaic properties of metal halide perovskites, including high carrier motilities, low carrier recombination rates, and the tunable spectral absorption range are attributed to the unique electronic properties of these materials. While DFT provides reliable structures and stabilities of perovskites, it performs poorly in electronic structure prediction. The relativistic GW approximation has been demonstrated to be able to capture electronic structure accurately, but at an extremely high computational cost. Here we report efficient and accurate band gap calculations of halide metal perovskites by using the approximate quasiparticle DFT-1/2 method. Using $AMX_3$ (A = $CH_3NH_3$, $CH_2NHCH_2$, Cs; M = Pb, Sn, X=I, Br, Cl) as demonstration, the influence of the crystal structure (cubic, tetragonal or orthorhombic), variation of ions (different A, M and X) and relativistic effects on the electronic structure are systematically studied and compared with experimental results. Our results show that the DFT-1/2 method yields accurate band gaps with the precision of the GW method with no more computational cost than standard DFT. This opens the possibility of accurate electronic structure prediction of sophisticated halide perovskite structures and new materials design for lead-free materials.

## Introduction:

Hybrid organic-inorganic lead halide perovskites have become the jewel in the crown in the field of photovoltaics since the pioneering work by Kojima *et al.* and Im *et al.*[1, 2]. In just four years the power conversion efficiency of solar cells based on hybrid perovskites has rapidly increased from an initial promising value of 9%[3] to over 22%[4]. In addition to the outstanding performance, simple and cost-effective fabrication, i.e. solution-processing techniques, render them one of the most desirable and most studied semiconductor materials in the field of photovoltaics[5]. Their extraordinary properties, including high carrier mobilities, low carrier recombination rates, and the tunable spectral absorption range are attributed to the unique electronic properties of these materials[5-7]. Thanks to the enormous interest from the scientific community, the field of hybrid perovskites has quickly expanded in terms of types of materials by substituting one or more of the organic or inorganic ions in one of the most studied perovskites, methylammonium lead iodide ($MAPbI_3$), to obtain the metal halide perovskites $AMX_3$: (A = Cs, $CH_3NH_3$, hereafter MA, $CH_2NHCH_2$, hereafter FA; M = Sn, Pb; X = I, Br, Cl).

Despite the rapid progress made in the last few years in terms of the conversion efficiency, the understanding of the fundamental electronic properties of $AMX_3$ perovskites is rather limited.

This is especially true for realistic structures in working devices, which often are compounds with mixtures of more than one organic cation, metal cation or halide anion with sophisticated surfaces and interfaces[1-7]. The challenges to fundamental understanding of the electronic structure of $AMX_3$ perovskites stem from their rich chemical and physical properties and the interplay of these properties[6, 7]. In this context, a first-principles computational approach capable of reliably calculating the materials properties (electronic and thermodynamic) is essential. While standard DFT provides reliable structures and stabilities of perovskites, it severely underestimates the band gaps[8-10]. The relativistic GW approximation has been demonstrated to be able to capture their electronic structure accurately but at an extremely high computational cost.

Several research groups[8-16] have successfully applied the GW method to predicting electronic structures of $AMX_3$ perovskites. Even *et al.*[8] have identified the importance of a giant spin-orbit coupling (SOC) effect (about 1.0 eV) in the lead iodide perovskites, acting mainly on the conduction band (mainly consisting of Pb states). Taking SOC into account in a relativistic GW approach, Brivio et al[9] predicted an unconventional band dispersion relation and a Dresselhaus splitting at the band edges in pseudo-cubic MAPbI$_3$, which indicates a direct-indirect band gap character. This theoretical prediction was recently confirmed[6, 7] by two groups of researchers independently, revealing the mechanism behind the slow charge recombination in these materials. Umari *et al.*[10] studied the substitution of Pb by Sn and found the SOC effect on Sn perovskites to be much smaller (about 0.4 eV). The same group of researchers also looked at the substitution of halide ions in MAPbX$_3$ [11] and nicely reproduced experimental findings in optical behavior, such as an increase of the band gap when moving from I to Cl. A few other theoretical studies at GW level include the investigation of polar phonons[12], crystal structure effects and phase transitions[13, 14], exciton binding energies[15], and band gap trends[16] in $AMX_3$. The above mentioned GW studies have been very important in understanding the chemistry and physics of these materials and providing materials design inspirations.

However, the GW calculations of realistic structures of metal halide perovskites remain challenging because of i) spin-orbit coupling effects ii) the complex structures and phase transitions in these materials iii) the necessary compromise between the size of the system studied and the extremely high computational cost. This leads to the quest for an accurate and cost effective theoretical framework for electronic structure calculations of realistic $AMX_3$ structures and possibly the A/M/X mixed compounds. Here, we apply a recently developed approximate quasiparticle method, namely the DFT-1/2 method[17, 18], which allows us to accurately model the band gaps of $AMX_3$ perovskites with minimal computational cost. The DFT-1/2 calculated electronic properties of $AMX_3$ perovskites are compared with those calculated with the GW method and with experimental data. Trends in the interplay of geometrical properties and electronic structure properties are analyzed systematically. Our results indicate that the DFT-1/2 method yields band gaps with a GW precision, but with a computational cost similar to standard DFT, opening the way to the study of sophisticated structures and new materials design.

**Computational methods and structural models:**

The initial structure optimizations are performed using DFT within the local density approximation (LDA) [19] as implemented in the Vienna *ab-initio* simulation package (VASP) [20]. The exchange-correlation (XC) functional is used as parameterized by Perdew and Zunger[21].

The outermost $s$, $p$, and $d$ (in the case of Pb and Sn) electrons are treated as valence electrons whose interactions with the remaining ions is modeled by pseudopotentials generated within the projector-augmented wave (PAW) method[22]. Fig. 1 shows the crystal structures and unit cells used in the DFT calculations. Taking MAPbI$_3$ as an example, unit cells with 12, 48, and (also) 48 atoms are used for the case of cubic, tetragonal, and orthorhombic crystal structures, respectively. During the optimization, the positions of the atoms, and the shape and volume of the unit cell are all allowed to relax. An energy cutoff of 500 eV and 10×10×10, 8×8×6, and 6×8×6 $k$-point meshes (for cubic, tetragonal and orthorhombic structures, respectively) are used to achieve energy and force convergence of 0.1 meV and 2 meV/Å, respectively.

DFT-LDA slightly underestimates the lattice parameters of the AMX$_3$ structures by about 1% to 3.5% depending on the composition of the compound and its structure. Usually, such an underestimation of lattice parameters leads to a small overestimation of the band gaps, e.g., by about 70-150 meV for group III/V semiconductors[23], taking volume deformation potentials into account. However, the opposite is found for metal halide perovskites, namely, an underestimation of the band gaps. The deviations of the band gaps (compared to those with experimental lattice constants) are in the range of 100 meV to 250 meV when using LDA-optimized lattice constants. To be consistent, all the electronic structure calculations were performed with corrected lattice parameters by expanding the lattice parameters proportionally (to match experimental volume of the cells) while keeping the LDA-optimized shape of the cells. This procedure keeps the *ab-initio* aspects of our approach without compromising accuracy. As an example, a band gap difference of 0.04 eV was found for MAPbI$_3$ when using this procedure (DFT-1/2 band gaps of 1.84 eV *vs* 1.88 eV).

The subsequent electronic structure calculations were performed using the DFT-1/2 method. The DFT-1/2 method stems from Slater's proposal of an approximation for the excitation energy, a transition state method[24, 25], to reduce the band gap inaccuracy by introducing a half-electron/half-hole occupation. Ferreira *et al.*[17] extended the method to modern DFT and particularly to solid-state systems, by assuming that the excited electron in the conduction band of a semiconductor usually occupies Bloch-like states with nearly vanishing self-energy, while the hole left in the valence band is localized with a finite self-energy. The self-energy of the hole was corrected by modifying the corresponding pseudopotentials of the atoms (in real space) by removing half an electron from the orbitals that contribute to the top of the valence band[17, 18]. The DFT-1/2 method has been demonstrated to be a successful approximate quasiparticle method for a large range of materials, including group III/V semiconductors, intermediate and large band gap nitrides, oxides, and other materials[17-18]. Fortunately, the computational effort is the same as for standard DFT, with a straightforward inclusion of SOC when coupled with VASP[20].

In this paper, we have applied the DFT-1/2 method (the LDA-1/2 version of Ref. 17) by using an occupation number of eight and proper cutoff radii (CUT) for the M cations and X anions in the AMX$_3$ compounds (Table 1). When determining these parameters, we have followed three rules: (i) the parameter CUT is chosen so as to maximize the band gap, ii) this parameter is transferable to different chemical environments in the family of metal halide perovskites, iii) the corrections of the A cations were not included mainly because A cations were demonstrated to not influence the band gap and band edges of metal halide perovskites. A schematic illustration of the application procedure of the DFT-1/2 method included in Figure 1S in the Supplementary Material. To validate the application of the DFT-1/2 method, a direct comparison of the band gaps obtained using the DFT-1/2 method with GW band gaps calculated by Brivio *et al.*[9] (MAPbI$_3$, FASnI$_3$) and Mosconi *et al.*[12] (AMX$_3$: A=MA/FA, M=Pb/Sn, X=I, Br, Cl) are shown in Table 1S of the Supplementary Material. The agreement between the DFT-1/2 and GW band gaps is excellent, with a maximum difference of less than 0.2 eV.

**Table 1** Values of the parameter CUT (in atomic units) for the half ionized orbitals used within the DFT-1/2 electronic structure calculations.

| Atom | CUT (a.u.) | Half-ionized orbital |
|------|-----------|---------------------|
| Pb   | 2.18      | d                   |
| Sn   | 2.30      | p                   |
| Cl   | 3.12      | p                   |
| Br   | 3.34      | p                   |
| I    | 3.76      | p                   |

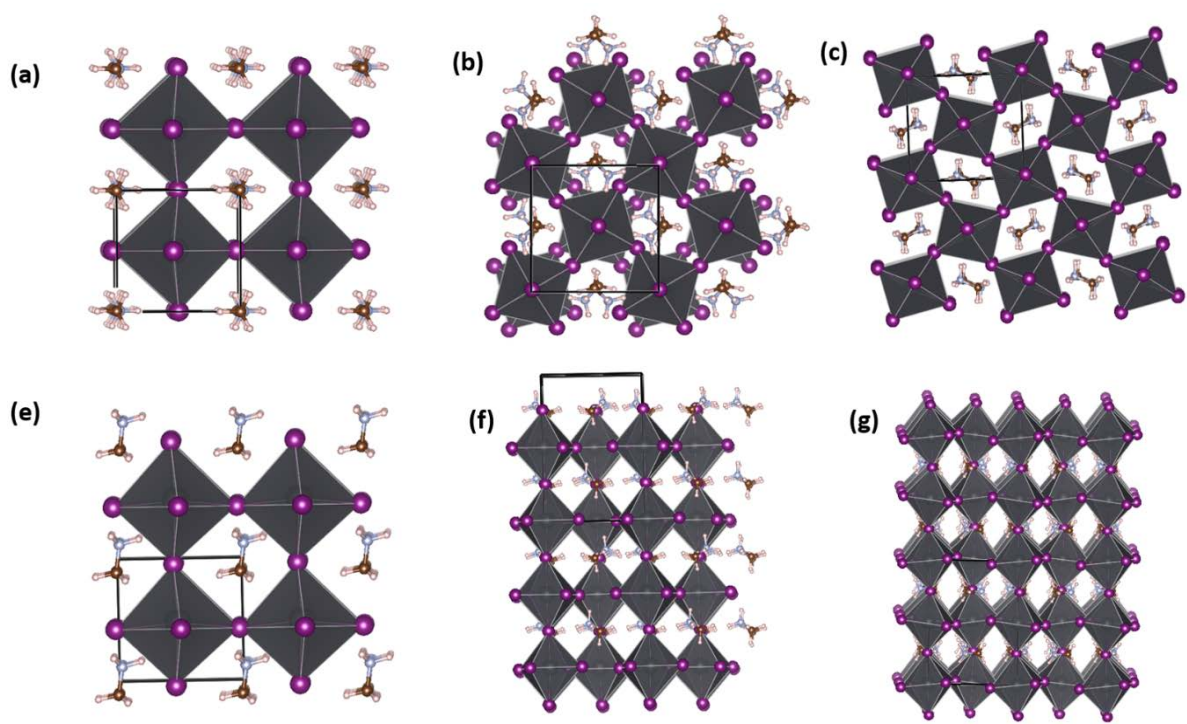

**Figure 1.** Top (a, b, and c) and side (e, f, and g) views of various structures of metal halide perovskites, taking MAPbI$_3$ as an example (visualization by VESTA[a]). From left to right: cubic, tetragonal and orthorhombic structures. The unit cell used in the DFT calculations are indicated by black lines.

[a] Ref 26.

**Results and discussion**

**Comparison of DFT and DFT-1/2 applied to pseudo-cubic MAPbI₃**

To compare the performance of DFT-1/2 with standard DFT in electronic structure calculations, we choose as an example the most studied metal halide perovskite MAPbI$_3$ with a pseudo-cubic structure. It is shown in Fig. 2(a) that DFT (LDA) indeed underestimates the band gap severely, which is found to be 1.47 eV and 0.39 eV without and with SOC, respectively. The tendency of DFT to underestimate band gaps is traditionally associated with the energy cost to excite electron-hole pair. Thanks to the self-energy correction, DFT-1/2 (LDA-1/2) performs extremely well with band gaps of 2.77 eV and 1.68 eV without and with SOC, respectively; see Fig. 2(b). The latter value is only 0.01 eV larger than the value of 1.67 eV calculated by Brivio *et al.*[9] using the GW approach. Both values are slightly higher than the experimental value of about 1.60 eV[28-29].

As compared to DFT, both the valence band (VB) and conduction band (CB) calculated with DFT-1/2 shifts downwards and upwards in energy, resulting in a band gap widening of about 1.30 eV both with and without SOC. (see Supplementary Material) In addition, the inclusion of SOC also leads to an indirect gap of 1.64 eV between the highest VB state and lowest CB state, which is 0.04 eV lower than the direct band gap. In comparison to other theoretical predictions of 20 meV of this difference using the HSE06 hybrid fuctional[27], and 75 meV[7] using self-consistent GW, our value is the close to the very recent experimental finding of an activation energy of 47 meV[6]. It should be noted that the direct-indirect character of the band gap in MAPbI$_3$ is also found by us in MAPbBr$_3$ and MAPbCl$_3$ for a pseudo-cubic structure, with energy differences of about 0.03 eV (not shown).

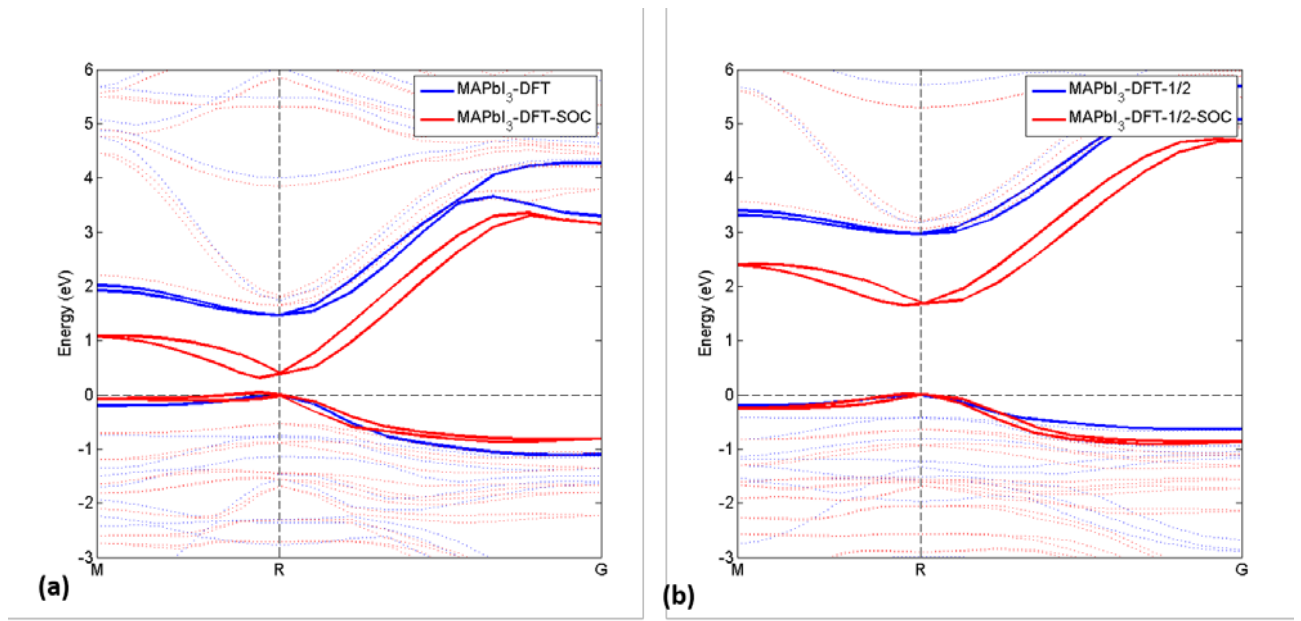

Figure 2. Comparison between calculated band structure of MAPbI$_3$ with (a) DFT (LDA) and (b) DFT-1/2 (LDA-1/2) for a pseudo-cubic structure with and without SOC. The highest-lying valence and lowest-lying conduction bands are highlighted as thick solid lines. The energy zero is set in both cases at the highest occupied state. DFT-LDA-optimized lattice parameters (8.542, 8.555, and 12.596) were used.

**Table 2.** Summary of experimental and DFT optimized lattice constants (in Å), band gap energies (eV) obtained with the DFT-1/2 method with and without SOC, compared with relativistic GW and experimental results. α, β, and γ denote, respectively, (pseudo-) cubic, tetragonal and orthorhombic structures of AMX$_3$. The details of the DFT-optimized crystal structures are given in the Supplementary Material.

| materials | optimized lattices constants [a] | experimental lattice constants | DFT-1/2 | DFT-1/2+SOC | Expt. | GW+SOC |
|---|---|---|---|---|---|---|
| α-MAPbI$_3$ | 6.172, 6.149, 6.218 (2.5%) | 6.328[d] | 2.96 | 1.81 | 1.60-1.61,[A-C], 1.65 [D,] 1.68[C] | 1.31-1.73 [M,N,O,P,Q] |
| β-MAPbI$_3$ | 8.542, 8.555, 12.596 (2.5%) | 8.86, 8.86, 12.66[d] | 2.84 | 1.84 | | |
| α-MAPbBr$_3$ | 5.539, 5.505, 5.644 (2%) | 5.675[d] | 3.49 | 2.40 | 2.33-2.35[D,E] | 2.34[O], 2.56[Q], 2.83[R] |
| α-MAPbCl$_3$ | 5.822, 5.802, 5.869 (1.2%) | 5.901[d] | 4.16 | 3.09 | 2.88-3.13[D-G] | 3.07[O], 3.46 [Q], 3.59[R] |
| α-FAPbI$_3$ | 6.321, 6.149, 6.216 (2%) | 6.362[e] | 2.47 | 1.38 | - | - |
| β-FAPbI$_3$ | 8.843, 8.843, 12.413 (2%) | - | 2.77 | 1.54 | 1.41-1.47[H,I] | 1.38[M]1.48[O] |
| α-CsPbI$_3$ | 6.135 (2%) | 6.289[f] | 2.54 | 1.44 | 1.74[J] | 1.62[R] |
| γ-CsPbI$_3$ | 8.959, 12.222, 7.933 (2%) | - | 3.05 | 2.00 | - | - |
| α-FASnI$_3$ | 6.220, 6.053, 6.133 (3.5%) | - | 1.40 | 1.10 | - | - |
| γ-FASnI$_3$ | 8.566, 12.097, 8.716 (3.5%) [b] | 8.930, 6.309, 9.062[g] | 1.54 | 1.23 | 1.41[K] | 1.27[O] |
| α-CsSnI$_3$ | 6.033 (3%) | 6.219[h] | 1.25 | 0.82 | - | 0.60[S], 1.01[T] |
| γ-CsSnI$_3$ | 8.649, 12.073, 8.190 (3%) | 8.688, 12.378, 8.643[h] | 1.71 | 1.34 | 1.30[L] | 1.30[P] |
| α-MASnI$_3$ | 6.073, 6.063, 6.133 (2.5%) | 6.230, 6.230, 6.232[i] | 1.29 | 0.91 | - | 1.03,[O] 0.89[R] |
| β-MASnI$_3$ | 8.432, 8.442, 12.377 (2.5%) | 8.758, 8.758, 12.429[i] | 1.52 | 1.14 | 1.21[U] | 1.10[N] |
| α-MASnCl$_3$ | 5.496, 5.401, 5.564 (2.5%) | 5.760[j] | 2.63 | 2.32 | - | 2.30[R] |
| γ -MASnCl$_3$ | 7.236, 10.961, 8.098 (2.5%)[c] | 7.910, 5.726, 8.227[j] | 2.53 | 2.22 | - | - |

[a] Values in brackets are the expansion percentages used to match the DFT-optimized unit cell volumes to the experimental ones.

[b, c] Experimentally the triclinic structure is observed at room temperature. However, to be consistent with other systems, the orthorhombic unit cell was used in the DFT calculations (the lattice constant $b$ is doubled).

d Ref. 30, e Ref. 31, f Ref. 32, g Ref. 33, h Ref. 34, i Ref. 35, j Ref. 36,

A Ref. 28, B Ref. 29, C Ref. 35, D Ref. 36, E Ref. 37, F Ref. 38, G Ref. 39, H Ref. 40, I Ref. 41, J Ref. 42, K Ref. 43, L Ref. 44,

M Ref. 9, N Ref. 10, O Ref. 12, P Ref. 14, Q Ref. 11, R Ref. 16, S Ref. 13, T Ref. 15, U Ref. 45

### DFT-1/2 results for $AMX_3$ (A = MA, FA, Cs, M = Pb, Sn, X = I, Br, Cl)

We applied the DFT-1/2 method to calculate the electronic structure of nine $AMX_3$ perovskites with their preferred crystal structures at room temperature[30-36]: i) (pseudo-) cubic for $MAPbBr_3$[30], $MAPbCl_3$[30], and $MASnCl_3$[36] ii) tetragonal for $MAPbI_3$[30], $FAPbI_3$[31], and $MASnI_3$[35] iii) orthorhombic for $CsPbI_3$[32], $FASnI_3$[33], and $CsSnI_3$[34]. It is well recognized that the orthorhombic↔tetragonal↔cubic (size-dependent, temperature-dependent, composition-dependent) phase transitions in $AMX_3$ perovskites can significantly alter their optical as well as electrical properties and thus impact their applications. Therefore, we also considered the low-temperature or high-temperature crystal structure for some of the compounds. Table 2 and Fig. 3 summarize DFT-optimized lattice parameters and DFT-1/2 calculated band gaps and their comparison with experimental data and GW results from literature[38-45].

Despite the approximation made in the procedure of adapting to the experimental lattice parameters (see Section computational methods) and the use of relatively small unit cells of the structures, the DFT-1/2 band gaps are in good agreement with experimental and GW results. The maximum discrepancy is about 0.2 eV, with the majority of the band gaps slightly overestimated and a few slightly underestimated. Special attention needs to be paid to the case of $CsPbI_3$ due to the fact that the room-temperature crystal structure is uncertain, which is an indication of a metastable phase at the experimental conditions. Therefore, one has to be careful when comparing a band gap value obtained from theoretical calculations for a certain structure that is different in experiments. The relatively large overestimation of the band gap of $CsPbI_3$ can be understood by assuming in the experiment an intermediate configuration between a cubic and orthorhombic assembling structure. Indeed, the average band gap of cubic (α-) $CsPbI_3$ and orthorhombic (γ-) $CsPbI_3$ is 1.72 eV, which is only 0.02 eV lower than the experimental value of 1.74 eV.

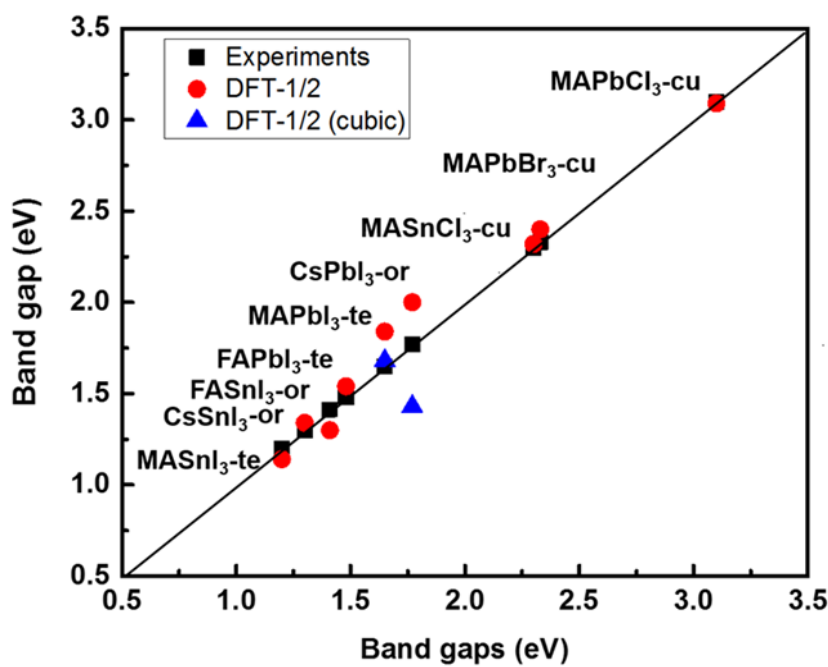

Figure 3 Comparison between experimental and DFT-1/2 band gaps of $AMX_3$ perovskites. Note: when the room-temperature crystal structure is uncertain (MAPbI$_3$ and CSPbI$_3$), the band gaps of the high-temperature crystal structure, i.e. the cubic phase, are also plotted. When no experimental result is available (MASnCl$_3$), the GW result is used.

A few general trends are observed from Fig. 3 and Table 2:

i)    For the same composition, the band gap decreases with increasing ordering in the crystal, namely orthorhombic > tetragonal > cubic. The magnitude of the differences is generally small for systems with MA, intermediate for FA, and large for Cs cations. Two extreme cases are MAPbI$_3$ with a difference of 0.03 eV (cubic *vs* tetragonal) and CsSnI$_3$ with a difference of 0.52 eV (cubic *vs* orthorhombic). The small band gap value of CsSnI$_3$ in a cubic structure was also predicted using the GW method[13, 15] (DFT-1/2: 0.82 eV, GW: 0.60 eV and 1.01 eV). This is probably related to the strong relaxation and reorientation of the organic cation (MA, FA) within the inorganic framework, resulting in pseudo-cubic structure (which is very close to a tetragonal structure), whereas the effect of Cs on the crystal structure is small, maintaining the perfect cubic structure.

ii)   In general, the band gap increases with an increase of electronegativity of the M and X ions: from I to Br to Cl and from Pb to Sn.

iii)  There is no general trend when considering the effect of the A cations. The effect of the A cations will be discussed in the following band structure analysis.

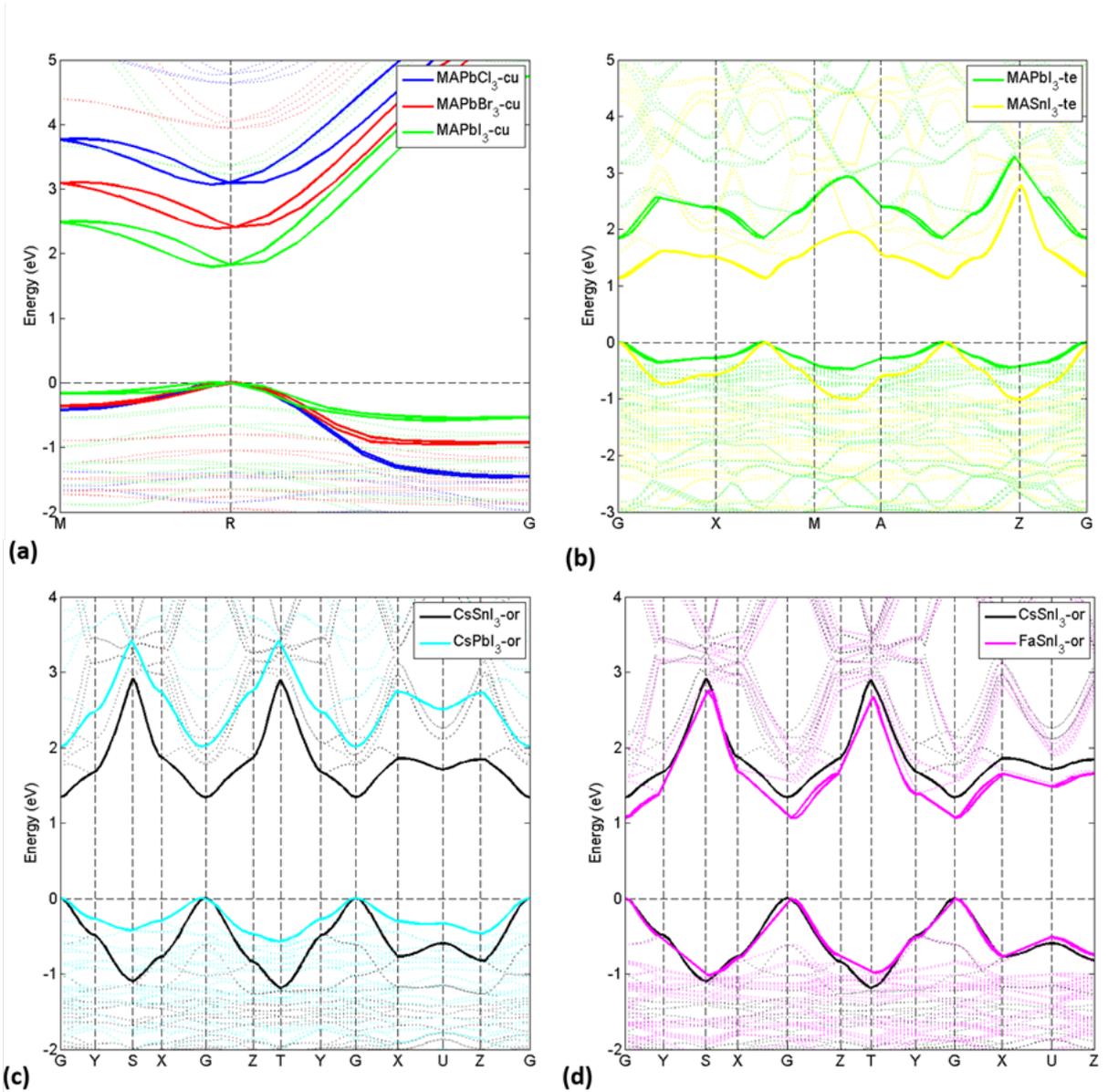

Figure 4. Calculated DFT-1/2 band structure (VBM and CBM highlighted as thick solid lines) for (a) MAPbX$_3$ (X = I, Br, Cl) in a pseudo-cubic structure, (b) MAMI$_3$ (M = Sn, Pb) with a tetragonal structure, (c) CsMI$_3$ (M = Sn, Pb) with an orthorhombic structure, (d) ASnI$_3$ (A = Cs, FA)with an orthorhombic structure. The energy zero is set in both cases at the highest occupied state.

**Electronic band structures and Density of States (DOS)**

To provide a better insight into the trends in the electronic structures of the nine AMX$_3$ perovskites, we plot the electronic band structures in Fig. 4. We group the compounds in terms of crystal structures and variations in only one of the ions (A/M/X): MAPbX$_3$ (X = I, Br, Cl) with a pseudo-cubic structure, MAMI$_3$ (M=Pb, Sn) in a tetragonal structure, CsMI$_3$ (M = Sn, Pb) with an orthorhombic structure, ASnI3 (A = Cs, FA) with an orthorhombic structure.

For MAPbX$_3$ (X = I, Br, Cl) with a pseudo-cubic structure, see Fig. 4(a), the band gaps and the trend in the increasing band gap when moving from I to Cl are in excellent agreement with experiments. The maximum discrepancy is less than 0.02 eV, which is within the inherent accuracy of the methodology. The agreement of our results with experiments, especially for the cases of MAPbBr₃ and MAPbCl₃, are better than in the work of Mosconi *et al.* [11] using the GW method, where sizable overestimations were found for a tetragonal structure. This again highlights the importance of the crystal structure of the studied materials when comparing with experiments. The common feature of all the simulated band structure are significant SOC effect with direct-indirect band gaps. The substitution of I by Br (Cl) has little influence on the band dispersion of the CB (consisting mainly of Pb 6p states) but leads to sizable changes in the VB (consisting mainly of halide *n*p states and to some extent of Pb 6s states). More dispersion in the VB is found when substituting I by Br (Cl), evidenced by the downshifting of the bands at the Γ point (Fig. 4(a)).

We pay special attention to two very important materials for photovoltaic applications, tetragonal MAPbI₃ and MASnI₃. Figs. 4(b) and 5 show the band structures and densities of states (DOS), respectively, together with a comparison with results from Umari[10]. As shown in Table 2, the band gap of 1.14 eV of MASnI₃ is in excellent agreement with the experimental value of 1.21 eV and the GW value of 1.10 eV. Similar to the case of MAPbX₃, the CB dispersion characteristics are almost unchanged. The substitution of Pb by Sn leads to a slight shift of the CB to higher energy by about 0.03 eV (see Fig. 5). On the contrary, the VB maximum is shifted upwards much more strongly by about 0.67 eV and the band dispersion increases (see the band dispersion between Γ and X, M and A, and at Z). These electronic structure differences between MAPbI₃ and MASnI₃ agree nicely with experimental findings in optical absorption spectra of a red-shift and an increased absorption intensity near the absorption onset[43]. In addition, our calculated DOS and band edge alignment is in good agreement with results using the GW method by Umari *et al.*[10]: 0.67 (0.03) *vs* 0.7 (0.2) eV for VB (CB) energy shifts.

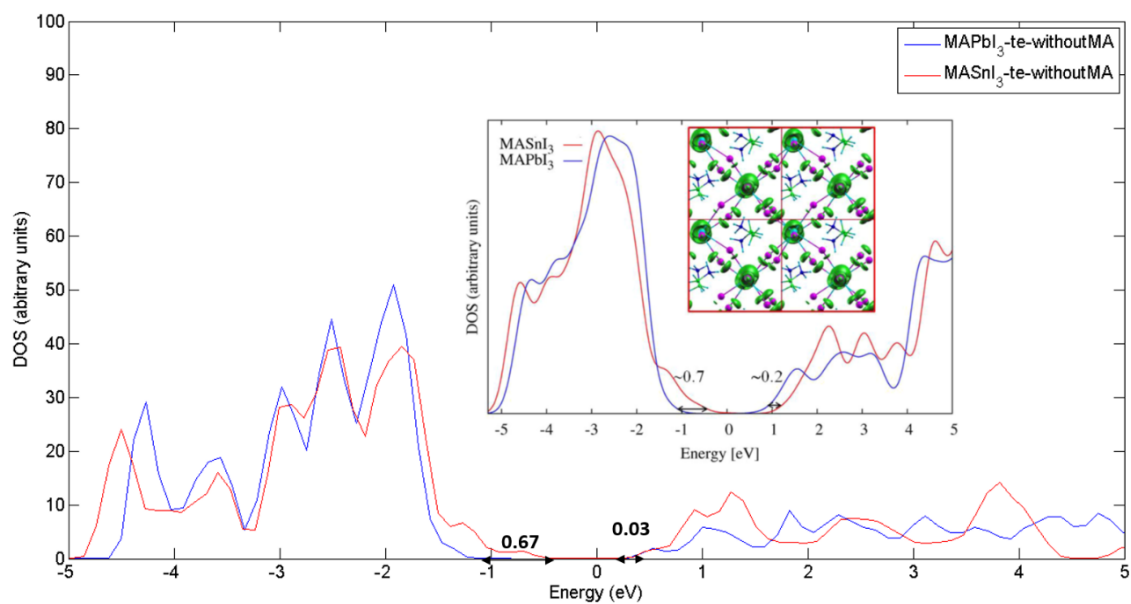

Figure 5 Electronic DOS for MAPbI₃ (blue) and MASnI₃ (red) calculated by DFT-1/2. The DOS peaks have been aligned at the localized I states at about -13 eV (Fig. 3S in the Supplementary Material). Inset: SOC-GW calculated DOS of MAPbI$_3$ and MASnI$_3$ from Umari *et al.*[10].

For CsMI$_3$ (M = Sn, Pb) with an orthorhombic structure (Fig.4(c)), a very similar trend is found as for MAMI$_3$ (M = Sn, Pb). The substitution of Pd by Sn leads to a smaller band gap. It should be noted that the DFT-1/2 band gaps of CsMI$_3$ are slightly larger than those of MAMI$_3$: 2.00 (1.30) eV *vs* 1.81 (1.14) eV for M = Pb (Sn). The difference in band gaps between CsMI$_3$ and MAMI$_3$ could be a result of variation in structures, unit cell sizes, and the detailed interactions of the A, M, and I ions. Common explanations for the difference are counteracting each other. For example, the volume of the CsPbI$_3$ unit cell is 7% smaller than that of the unit cell of MAPbI$_3$ (usually leading to a smaller band gap), whereas the degree of the tilting of the PbI$_6$ octahedral framework in CsPbI$_3$ is larger than that for MAPbI$_3$ (usually leading to a larger band gap). While both the volume and the tilting of the octahedral framework of CsSnI$_3$ are slightly smaller than those of the MASnI$_3$by less than 2% (both usually leading to a smaller band gap), the calculated band gapof CsSnI$_3$ is in fact larger than that of MASnI$_3$. This unexpected result may be explained by taking into account the different chemistry and interactions of Cs and MA with the inorganic matrix of AMX$_3$.

As expected, for ASnI$_3$ (A = Cs, FA) with an orthorhombic structure (Fig.4 (d)), the change of cation from Cs to FA yields almost no change in band dispersion and a slight band gap decrease. This is in agreement with one of the commonly accepted features of metal halide perovskites: the A cation only weakly interacts with the inorganic matrix, with little electronic contribution at the band edges, and mainly plays a role in changing the volume of the lattice. The band gap differences between FASnI$_3$ and MASnI$_3$ are again due to combinations of differences in unit cell volume (FASnI$_3$ > CsSnI$_3$), tilting of the octahedral framework (FASnI$_3$ > CsSnI$_3$), and chemical bonding characteristics. To summarize, our analysis emphasizes the fact that the electronic structure of AMX$_3$ perovskites is a result of the interplay of several properties: chemical environment, crystal structure, unit cell size, and thermodynamics of the materials. Therefore, an accurate electronic structure description of AMX3 perovskites requires not only a reliable *ab-initio* method but also a proper understanding of the interplay of several chemical and physical properties of the materials.

In conclusion, we have applied an approximate quasiparticle method, the DFT-1/2 method, to electronic structure calculations of AMX$_3$ perovskites (A = CH$_3$NH$_3$, CH$_2$NHCH$_2$, Cs; M = Pb, Sn, X = I, Br, Cl) with several crystal structures. Our results show that the DFT-1/2 method yields accurate band gaps showing very good agreement with experimental findings and results from the computationally much more demanding GW method. Nevertheless, the computational cost is comparable to that of a standard DFT method. In addition, the trends in electronic structure properties were identified by varying only one of the three A/M/X ions. In general, the band gaps increase with an increase in electronegativity of the M and X ion, and increase with a decrease of the symmetry of the crystal structures: cubic < tetragonal < orthorhombic. There is no single applicable rule governing the trend in band gaps when only varying the A cation. Here, the interplay of several aspects, such as unit cell size, crystal structure, chemical bonding, and thermodynamics of the material is decisive. Our work demonstrates the success of the DFT-1/2 method in predicting the electronic structure of the AMX$_3$ perovskites with minimal computational cost and opens the pathway towards studying large and complex structures in working devices and efficiently designing new metal halide perovskite materials for solar cell applications.

**References:**

1. Kojima, A., Teshima, K., Shirai, Y., & Miyasaka, T. Organometal halide perovskites as visible-light sensitizers for photovoltaic cells. *Journal of the American Chemical Society* **131**, 6050-6051, doi: 10.1021/ja809598r (2009).

2. Im, J. H., Lee, C. R., Lee, J. W., Park, S. W., & Park, N. G. 6.5% efficient perovskite quantum-dot-sensitized solar cell. *Nanoscale* **3**, 4088-4093, doi: 10.1039/C1NR10867K (2011).

3. Lee, M. M., Teuscher, J., Miyasaka, T., Murakami, T. N., & Snaith, H. J. Efficient hybrid solar cells based on meso-superstructured organometal halide perovskites. *Science* **338**, 643-647, doi:10.1126/science.1228604 (2012).

4. Saliba, M., Matsui, T., Domanski, K., Seo, J. Y., Ummadisingu, A., Zakeeruddin, S. M. *et al*. Incorporation of rubidium cations into perovskite solar cells improves photovoltaic performance. *Science* **354**, 206-209, doi:10.1126/science.aah5557 (2016).

5. Moser, J. E. Perovskite photovoltaics: Slow recombination unveiled. *Nature Materials* **16**, 4-6, doi: 10.1038/nmat4796 (2017).

6. Hutter, E. M., Gélvez-Rueda, M. C., Osherov, A., Bulović, V., Grozema, F. C., Stranks, S. D., & Savenije, T. J. Direct–indirect character of the bandgap in methylammonium lead iodide perovskite. *Nature Materials* **16**, 115-120, doi:10.1038/nmat4765 (2017).

7. Azarhoosh, P., McKechnie, S., Frost, J. M., Walsh, A., & Van Schilfgaarde, M. (2016). Research Update: Relativistic origin of slow electron-hole recombination in hybrid halide perovskite solar cells. *APL Materials* **4**, 091501, doi: http://dx.doi.org/10.1063/1.4955028 (2016).

8. Even, J., Pedesseau, L., Jancu, J. M., & Katan, C. Importance of spin–orbit coupling in hybrid organic/inorganic perovskites for photovoltaic applications. *The Journal of Physical Chemistry Letters* **4**, 2999-3005, doi: 10.1021/jz401532q. (2013).

9. Brivio, F., Butler, K. T., Walsh, A., & Van Schilfgaarde, M. Relativistic quasiparticle self-consistent electronic structure of hybrid halide perovskite photovoltaic absorbers. *Physical Review B* **89**, 155204, doi:10.1103/PhysRevB.89.155204 (2014).

10. Umari, P., Mosconi, E., & De Angelis, F. Relativistic GW calculations on CH3NH3PbI3 and CH3NH3SnI3 perovskites for solar cell applications. *Scientific reports* **4**, 4467, doi: 10.1038/srep04467 (2014).

11. Mosconi, E., Umari, P., & De Angelis, F. Electronic and optical properties of MAPbX 3 perovskites (X= I, Br, Cl): a unified DFT and GW theoretical analysis. *Physical Chemistry Chemical Physics* **18**, 27158-27164, doi:10.1039/C6CP03969C (2016).

12. Bokdam, M., Sander, T., Stroppa, A., Picozzi, S., Sarma, D. D., Franchini, C., & Kresse, G. Role of Polar Phonons in the Photo Excited State of Metal Halide Perovskites. *Scientific Reports* **6**, doi:10.1038/srep28618 (2016).

13. Even, J., Pedesseau, L., Jancu, J. M., & Katan, C. DFT and k· p modelling of the phase transitions of lead and tin halide perovskites for photovoltaic cells. *physica status solidi (RRL)-Rapid Research Letters* **8**, 31-35, doi:10.1002/pssr.201308183 (2014).

14. Gao, W., Gao, X., Abtew, T. A., Sun, Y. Y., Zhang, S., & Zhang, P. Quasiparticle band gap of organic-inorganic hybrid perovskites: Crystal structure, spin-orbit coupling, and self-energy effects. *Physical Review B* **93**, 085202, doi:10.1103/PhysRevB.93.0852 02 (2016).

15. Huang, L. Y., & Lambrecht, W. R. Electronic band structure, phonons, and exciton binding energies of halide perovskites CsSnCl 3, CsSnBr 3, and CsSnI 3. *Physical Review B* **88**, 165203, doi:10.1103/PhysRevB.88.165203 (2013).

16. Castelli, I. E., García-Lastra, J. M., Thygesen, K. S., & Jacobsen, K. W. Bandgap calculations and trends of organometal halide perovskites. *APL Materials* **2**, 081514, doi: 10.1063/1.4893495 (2014).

17. Ferreira, L. G., Marques, M., & Teles, L. K. Approximation to density functional theory for the calculation of band gaps of semiconductors. *Physical Review B* **78**, 125116, doi:10.1103/PhysRevB.78.125116 (2008).

18. Ferreira, L. G., Marques, M., & Teles, L. K. Slater half-occupation technique revisited: the LDA-1/2 and GGA-1/2 approaches for atomic ionization energies and band gaps in semiconductors. *AIP Advances* **1**, 032119, doi: 10.1063/1.3624562 (2011).

19. Kohn, W., & Sham, L. J. Self-consistent equations including exchange and correlation effects. *Physical review* **140**, A1133, doi:10.1103/PhysRev.140.A1133 (1965).

20. Kresse, G., & Furthmüller, J. Efficiency of ab-initio total energy calculations for metals and semiconductors using a plane-wave basis set. *Computational Materials Science* **6**, 15-50, doi:10.1016/0927-0256(96)00008-0 (1996).

21. Perdew, J. P., & Zunger, A. Self-interaction correction to density-functional approximations for many-electron systems. *Physical Review B* **23**, 5048, doi:10.1103/PhysRevB.23.5048 (1981).

22. Kresse, G., & Joubert, D. From ultrasoft pseudopotentials to the projector augmented-wave method. *Physical Review B* **59**, 1758, doi:10.1103/PhysRevB.59.1758 (1999).

23. Belabbes, A., Panse, C., Furthmüller, J., & Bechstedt, F. Electronic bands of III-V semiconductor polytypes and their alignment. *Physical Review B* **86**, 075208, doi:10.1103/PhysRevB.86.075208 (2012).

24. Slater, J. C. Statistical exchange-correlation in the self-consistent field. *Advances in quantum chemistry* **6**, 1-92, doi:10.1016/S0065-3276(08)60541-9 (1972).

25. Slater, J. C., & Johnson, K. H. Self-consistent-field X α cluster method for polyatomic molecules and solids. *Physical Review B* **5**, 844, doi:10.1103/PhysRevB.5.844 (1972).

26. Momma, K., & Izumi, F. VESTA 3 for three-dimensional visualization of crystal, volumetric and morphology data. *Journal of Applied Crystallography* **44**, 1272-1276, doi:10.1107/S0021889811038970 (2011).

27. Motta, C., El-Mellouhi, F., Kais, S., Tabet, N., Alharbi, F., & Sanvito, S. Revealing the role of organic cations in hybrid halide perovskite CH3NH3PbI3. *Nature communications* **6**, doi: 10.1038/ncomms8026 (2015).

28. Yamada, Y., Nakamura, T., Endo, M., Wakamiya, A., & Kanemitsu, Y. Near-band-edge optical responses of solution-processed organic–inorganic hybrid perovskite CH3NH3PbI3 on mesoporous TiO2 electrodes. *Applied Physics Express* **7**, 032302, doi:10.7567/APEX.7.032302/meta (2014).

29. Chen, L. C., & Weng, C. Y. (2015). Optoelectronic Properties of MAPbI 3 Perovskite/Titanium Dioxide Heterostructures on Porous Silicon Substrates for Cyan Sensor Applications. *Nanoscale research letters* **10**, 404, doi: 10.1186/s11671-015-1114-x (2015).

30. Poglitsch, A., & Weber, D. Dynamic disorder in methylammoniumtrihalogenoplumbates (II) observed by millimeter-wave spectroscopy. *The Journal of chemical physics* **87**, 6373-6378, doi:10.1063/1.453467 (1987).

31. Weller, M. T., Weber, O. J., Frost, J. M., & Walsh, A. Cubic perovskite structure of black formamidinium lead iodide, α-[HC (NH2) 2] PbI3, at 298 K. *The Journal of Physical Chemistry Letters* **6**, 3209-3212, doi:10.1021/acs.jpclett.5b01432 (2015).

32. Trots, D. M., & Myagkota, S. V. High-temperature structural evolution of caesium and rubidium triiodoplumbates. *Journal of Physics and Chemistry of Solids* **69**, 2520-2526, doi:10.1016/j.jpcs.2008.05.007 (2008).

33. Koh, T.M., Krishnamoorthy, T., Yantara, N., Shi, C., Leong, W.L., Boix, P.P., Grimsdale, A.C., Mhaisalkar, S.G. & Mathews, N. Formamidinium tin-based perovskite with low E g for photovoltaic applications. *Journal of Materials Chemistry A* **3**, 14996-15000, doi: 10.1039/C5TA00190K (2015).

34. Yamada, K., Funabiki, S., Horimoto, H., Matsui, T., Okuda, T., & Ichiba, S. Structural phase transitions of the polymorphs of CsSnI3 by means of rietveld analysis of the X-ray diffraction. *Chemistry Letters* **20**, 801-804, doi:10.1246/cl.1991.801 (1991).

35 Phuong, L.Q., Yamada, Y, Nagai, M., Maruyama, N, Wakamiya, A., Kanemitsu, Y. Free Carriers versus Excitons in CH3NH3PbI3 Perovskite Thin Films at Low Temperatures: Charge Transfer from the Orthorhombic Phase to the Tetragonal Phase. J. Phys. Chem. Lett. 2016, 7, 2316−2321. J. Phys. Chem. Lett. 2016, 7, 2316−2321, http://dx.doi.org/10.1021/acs.jpclett.6b00781.

36. Papavassiliou, G. & Koutselas, I. Structural, optical and related properties of some natural three- and lower-dimensional semiconductor systems. *Synthetic Metals* **71,** 1713–1714 (1995).

37. Kitazawa, N., Watanabe, Y., & Nakamura, Y. Optical properties of CH3NH3PbX3 (X= halogen) and their mixed-halide crystals. *Journal of materials science* **37**, 3585-3587, doi: 10.1023/A: 1016584519829 (2002).

38. Ryu, S., Noh, J. H., Jeon, N. J., Kim, Y. C., Yang, W. S., Seo, J., & Seok, S. I. Voltage output of efficient perovskite solar cells with high open-circuit voltage and fill factor. *Energy & Environmental Science* **7**, 2614-2618, doi: 10.1039/C4EE00762J (2014).

39. Papavassiliou, G. C., & Koutselas, I. B. Structural, optical and related properties of some natural three-and lower-dimensional semiconductor systems. *Synthetic Metals* **71**, 1713-1714, doi:10.1016/0379-6779(94)03017-Z (1995).

40. Dimesso, L., Quintilla, A., Kim, Y. M., Lemmer, U. & Jaegermann, W. Investigation of formamidinium and guanidinium lead tri-iodide powders as precursors for solar cells. *Materials Science and Engineering: B* **204**, 27-33, doi:10.1016/j.mseb.2015.11.006 (2016).

41. Koh, T. et al. Formamidinium-containing metal-halide: an alternative material for near-IR absorption perovskite solar cells. *The Journal of Physical Chemistry C* **118**, 16458-16462, doi: 10.1021/jp411112k (2013).

42. Zhumekenov, A.A., Saidaminov, M.I., Haque, M.A., Alarousu, E., Sarmah, S.P., Murali, B., Dursun, I., Miao, X.H., Abdelhady, A.L., Wu, T. and Mohammed, O.F. Formamidinium lead halide perovskite crystals with unprecedented long carrier dynamics and diffusion length. *ACS Energy Letters* **1**, 32-37, doi:10.1021/acsenergylett.6b00002 (2016).

43. McMeekin, D. P. et al. A mixed-cation lead mixed-halide perovskite absorber for tandem solar cells. *Science* **351**, 151-155, doi:10.1126/science.aad5845 (2016).

44. Protesescu, L., Yakunin, S., Bodnarchuk, M.I., Krieg, F., Caputo, R., Hendon, C.H., Yang, R.X., Walsh, A. & Kovalenko, M.V. Nanocrystals of cesium lead halide perovskites (CsPbX3, X= Cl, Br, and I): novel optoelectronic materials showing bright emission with wide color gamut. *Nano letters* **15**, 3692-3696, doi: 10.1021/nl5048779 (2015).

45. Stoumpos, C. C., Malliakas, C. D. & Kanatzidis, M. G. Semiconducting tin and lead iodide perovskites with organic cations: Phase transitions, high mobilities, and near-infrared photoluminescent properties. *Inorg. Chem.* **52,** 9019–9038 (2013)

**Acknowledgements**

S.X.Tao acknowledge funding by the Computational Sciences for Energy Research (CSER) tenure track program of Shell, NWO, and FOM (Project number15CST04-2).

The authors acknowledge Dr. Peter Klaver for technical support in the use of VASP and Dr. Ross Hatton, Dr. Selina Olthof, and Prof. René Janssen for useful discussions.

**Author Contributions**

S.X.T. and P.A.B. conceived the project; S.X.T implemented the DFT-1/2 method to calculate band structures, and performed all DFT calculations. X.C. analyzed the results and made the figures and tables under the guidance of S.X.T.. All authors contributed to the discussions and writing of the manuscript.

**Additional Information**

**Competing financial interests:** The authors declare no competing financial interests.

# Supplementary information to "Accurate and efficient band gap predictions of metal halide perovskites using the DFT-1/2 method: GW accuracy with DFT expense"

S. X. Tao[1*], X. Cao[1], P. A. Bobbert[1]

Table 1. Direct comparison of band gaps obtained using DFT-1/2 method with GW results from Brivio et al[1] (MAPbI$_3$, FASnI$_3$) and from Mosconi et al [2] (AMX$_3$: A=MA/FA, M=Pb/Sn, X=I, Br, Cl). Note the crystals structures of GW work were used directly for DFT-1/2 band gap calculations.

|  | DFT-1/2 *vs* GW Ref 1 | DFT-1/2 *vs* GW Ref 2 |
|---|---|---|
| MASnI$_3$ |  | 1.07 (1.03) |
| MAPbI$_3$ | 1.62(1.67) | 1.74 (1.67) |
| FAPbI$_3$ |  | 1.66 (1.48) |
| MAPbBr$_3$ |  | 2.40 (2.34) |
| MAPbCl$_3$ |  | 3.10 (3.07) |
| FaSnI$_3$ |  | 1.21(1.27) |

1. Brivio, F., Butler, K. T., Walsh, A., & Van Schilfgaarde, M. Relativistic quasiparticle self-consistent electronic structure of hybrid halide perovskite photovoltaic absorbers. Physical Review B 89, 155204, doi:10.1103/PhysRevB.89.155204 (2014).

2. Bokdam, M., Sander, T., Stroppa, A., Picozzi, S., Sarma, D. D., Franchini, C., & Kresse, G. Role of Polar Phonons in the Photo Excited State of Metal Halide Perovskites. Scientific Reports 6, doi: 10.1038/srep28618 (2016).

Figure 1. A schematic illustration of the procedure of the DFT-1/2 method in predicting band structure of the pseudo-cubic MAPbI₃ perovskites.

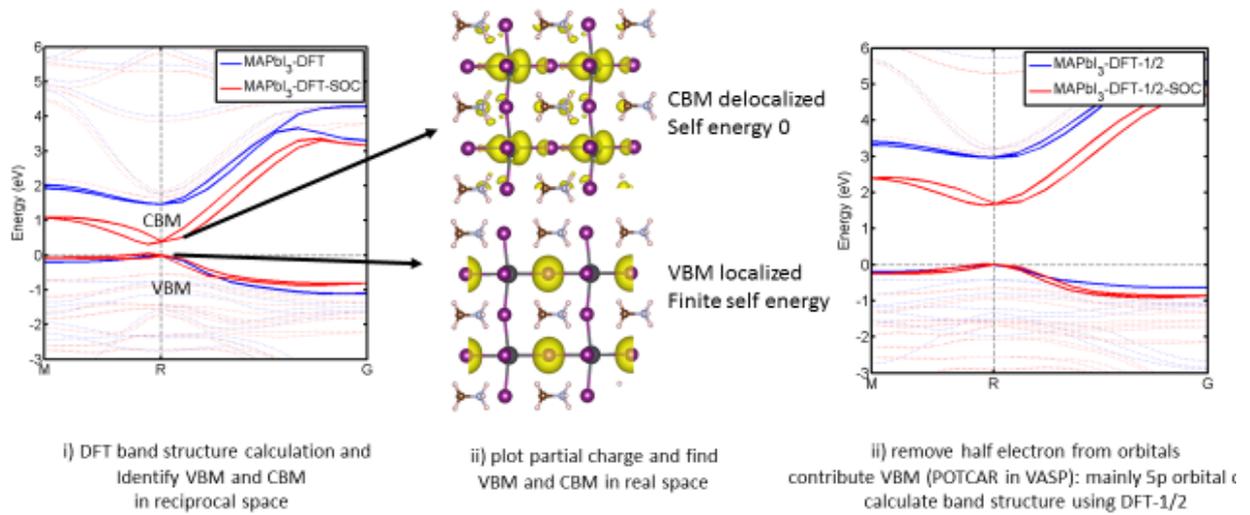

Figure 2 Band structure comparison of MAPbI₃ using DFT and DFT-1/2 without aligning the Valence band maximum to 0.

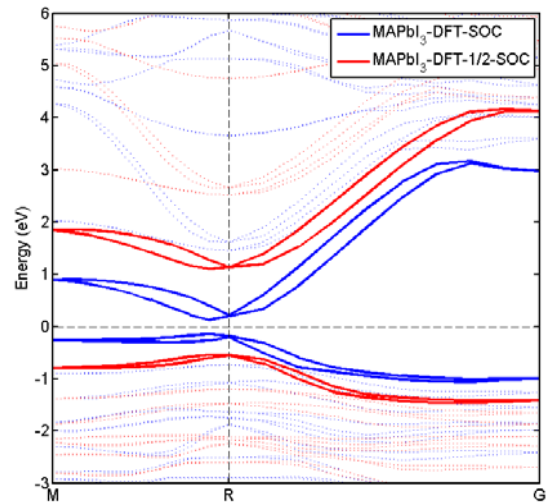

Figure 3 Electronic DOS for MAPbI₃ (blue) and MASnI₃ (red) calculated by DFT-1/2 (energy range of -25 eV to 10 eV). The DOS peaks have been aligned at the localized I states at about -13 eV.

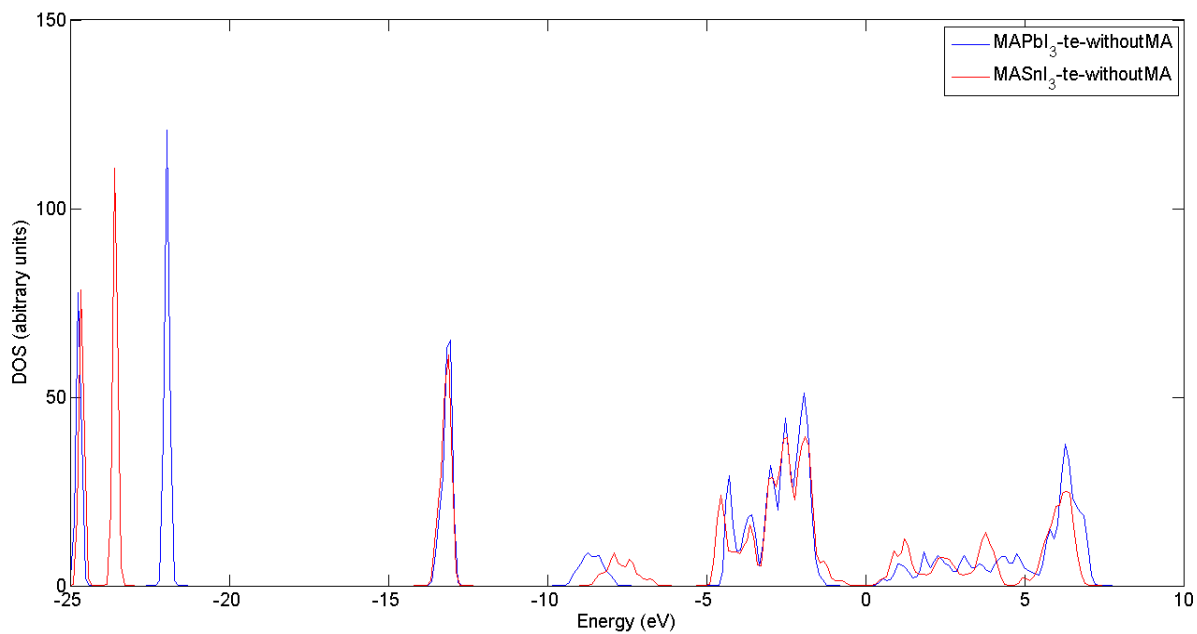

Optimized crystal structures formatted as VASP input file used in DFT-1/2 band structure calculations list in Table 2 in the manuscript:

-------------------------------------------------------------------------------------------------------

MAPbI3-cubic

  1.02500000000000

    6.1717394338931628    -0.0008531882374086    -0.0501585885085331

    -0.0008500076402037    6.1395863452849246    -0.0017851626838633

    -0.0503636493782260    -0.0017745042043623    6.2184346873496246

  C   N   H   Pb   I

   1   1   6   1   3

Direct

  0.9007168513900297  0.9999345777638595  0.9910526733883387

  0.1357170184255878  0.9996953788852068  0.0343372829410100

  0.8688585974588960  0.9996782355626976  0.8162393690239895

  0.8298756584273477  0.1470403438399472  0.0632866594003190

  0.8296216200505384  0.8531616513752454  0.0637072890186232

  0.2120804707301787  0.1378552707506628  0.9713724290568067

  0.2118653096297933  0.8613247821438677  0.9717113719033321

  0.1744096796530812  0.9999300129253115  0.1995764280873118

  0.4869999191442815  0.5002704598543701  0.4808893771358029

  0.4382965967767589  0.4999387017890840  0.9775135980187386

  0.4385244634437342  0.0002746974115269  0.5226928142736895

  0.9748428438697800  0.5001668976982074  0.4277616977520395

-----------------------------------------------------------------------------------------------------------------

MAPbI3-tetragonal

   1.0250000000000000

     8.5415512571730385    0.0018830583307183   -0.0071806817532417
     0.0085434065676154    8.5551944276888925   -0.0038934495258479
     0.1002173780274798    0.0764313213134684   12.5961575046758263
   C   N   H   Pb   I
   4   4   24   4   12

Direct

  0.0296868046819654  0.4976872446520417  0.2502455073755883
  0.5322070288555878  0.0458578332664885  0.2498911066309191
  0.0453407152028618  0.5328029454638070  0.7491792621135360
  0.4978644807673902  0.0302492564135761  0.7505469944809917
  0.9260752069455549  0.6079861523570500  0.1949687551622645
  0.4281071909336873  0.9335109319068167  0.1979147343360950
  0.9342638262333480  0.4270875187270917  0.6974564314387806
  0.6073524767287566  0.9254538024452330  0.6953367116349796
  0.6544776764267013  0.0197735953964298  0.2274330955993307
  0.5175286121221561  0.0355047437770750  0.3366653401111392
  0.5015625576087288  0.1649673116909796  0.2232611623828262
  0.4532970048051865  0.8175859618659089  0.2199526683332209
  0.4396342277609193  0.9402473279412575  0.1153810558696406
  0.3097226810141294  0.9546846482751121  0.2161429016602128
  0.9360572836629544  0.5957553163895071  0.1126626844912835
  0.9529121702683838  0.7248348997455167  0.2129068782896013
  0.8078069737922533  0.5890969820352439  0.2147485517468297
  0.1519218579776904  0.5208364272673549  0.2262680388177856
  0.9977344710413121  0.3772173580150593  0.2282052470981242
  0.0160455930475578  0.5129198026800168  0.3366542843050269
  0.9539627029102959  0.3096536221325792  0.7184552533675941

| | | |
|---|---|---|
| 0.9447657303213006 | 0.4341958889896702 | 0.6148805273227183 |
| 0.8175031573747376 | 0.4538946765757359 | 0.7169639554048430 |
| 0.1654364290027388 | 0.5003770852426257 | 0.7252548935440402 |
| 0.0315997698389694 | 0.5221628377916758 | 0.8360107277031972 |
| 0.0208442618777624 | 0.6539100737999064 | 0.7237487190827636 |
| 0.7246804507639126 | 0.9528024640925992 | 0.7118119539305496 |
| 0.5932355025331972 | 0.9330000169493289 | 0.6130736986300036 |
| 0.5889493051005630 | 0.8078446475102155 | 0.7165659578695482 |
| 0.5205548846131904 | 0.1516714700773463 | 0.7249874703239740 |
| 0.5147040005903492 | 0.0188179513111564 | 0.8369278800460833 |
| 0.3768113524687777 | 0.9977223310251375 | 0.7300765433937784 |
| 0.5092991392396584 | 0.5161104582173337 | 0.9888240789665872 |
| 0.5170377983743322 | 0.5101708882156970 | 0.4888165075324551 |
| 0.0296757111748249 | 0.0117638428962650 | 0.4898746508355032 |
| 0.0107165047973155 | 0.0295990521928147 | 0.9899093535600372 |
| 0.7003676003705834 | 0.8333652693744611 | 0.9949361905493603 |
| 0.3283229593061634 | 0.2051690134337534 | 0.9978877294952753 |
| 0.2051423522933149 | 0.7178159990791428 | 0.9991227281632789 |
| 0.8341064710660575 | 0.3470987376852506 | 0.9893416892492723 |
| 0.3471068455790700 | 0.8341797681447645 | 0.4895693405043957 |
| 0.7174528129028630 | 0.2055979374525450 | 0.4997599311188239 |
| 0.8339474806398144 | 0.7013754087403044 | 0.4951825901085911 |
| 0.2054387616907150 | 0.3289275027440368 | 0.4969014482428307 |
| 0.5221077370772633 | 0.5293229615830555 | 0.2401512535509411 |
| 0.5294315622363115 | 0.5221232972172203 | 0.7401440582814232 |
| 0.0091920552054177 | 0.0200349575165646 | 0.7415109482564759 |
| 0.0200060957733754 | 0.0092277656972470 | 0.2414863580874496 |

---------------------------------------------------------------------------------------------------------------

MAPbBr3-cubic

  1.01200000000000

   5.9199999999999999    0.0000000000000000    0.0000000000000000

   0.0000000000000000    5.9199999999999999    0.0000000000000000

   0.0000000000000000    0.0000000000000000    5.9199999999999999

  C   N   H   Pb   Br

   1   1   6   1   3

Direct

  0.8962732309267807  0.9999835575561917  0.9885225897978458

  0.1421229416578313  0.9998276714532324  0.0326818543098426

  0.8669576230289806  0.9997673821304502  0.8051964671801670

  0.8224554777663400  0.1524554610094526  0.0646689097381810

  0.8222031293734631  0.8478330826230547  0.0650649063082653

  0.2212941063647946  0.1424140526496274  0.9656588034540547

  0.2210818272207788  0.8569803126825306  0.9659456519655976

  0.1804492545013900  0.9999679128432390  0.2053791894684025

  0.4753317996463835  0.5000043839523016  0.4779234319137089

  0.4287771039399573  0.4998766652154316  0.9726919535286029

  0.4334819150903257  0.0000195562303063  0.5128482885866177

  0.9675529784829351  0.5001724086541017  0.4392056937486757

---------------------------------------------------------------------------------------------------------

MAPbCl3-cubic

  1.02000000000000

   5.5391045274592825    0.0003468956896641   -0.0010448343194384

   0.0003448753182629    5.5052137797524301    0.0001664298461292

  -0.0013785306998623    0.0001752310013020    5.6439202901432122

  C   N   H   Pb   Cl

  1   1   6   1   3

Direct

 0.8899586145506504 0.9999611127432928 0.9746996321370602

 0.1446557781237630 0.9998328693923924 0.0449578418890582

 0.8771793266459511 0.9998495446299316 0.7798749649489167

 0.8020933139467843 0.1645140612286227 0.0452580169852510

 0.8019401384725882 0.8356367022449902 0.0454669481036447

 0.2358990065205688 0.1540137764715652 0.9823668497221618

 0.2357756327953524 0.8455675023478406 0.9825229143105929

 0.1692202440177226 0.9999137203198316 0.2298624101125810

 0.4820195458734844 0.4999994259267098 0.4829510655592770

 0.4383591198597827 0.4998330362334400 0.9753494202562720

 0.4185616216959005 0.9999871121174451 0.5341565224073932

 0.9823190454974124 0.5001935833438793 0.4183211535678311

---------------------------------------------------------------------------------------------------------

FAPbI$_3$-cubic

  1.020000000000000

   6.3211144246449447   -0.0000000157951959    0.0000000000000000

   0.0000008075629942    6.1488880004277604    0.0000000000000000

   0.0000000000000000    0.0000000000000000    6.2156829632934336

  C   N   H   Pb   I

   1    2    5    1    3

Direct

  0.5000010130025601  0.5220356819570853  0.5000000000000000

  0.6833221224189714  0.4256937661911380  0.5000000000000000

  0.3166759105786099  0.4256937661860505  0.5000000000000000

  0.5000000000005471  0.7009747178519632  0.5000000000000000

  0.8157884239113073  0.5224359538161281  0.5000000000000000

  0.7038211913288990  0.2596793368997012  0.5000000000000000

  0.2961788086693744  0.2596793368993673  0.5000000000000000

  0.1842115760901422  0.5224359538145862  0.5000000000000000

  0.9999999999998934  0.9647834082093283  0.0000000000000000

  0.4999999999999218  0.9837789138322748  0.0000000000000000

  0.9999999999998010  0.4622583035517493  0.0000000000000000

  0.9999999999999645  0.9190658117906239  0.5000000000000000

FAPbI$_3$-tetragonal

  1.020000000000000

   8.8399587465740872   -0.0013088808757011   -0.0000046974807339

   -0.0013088639889776    8.8399578084821986   -0.0000001502236795

   -0.0000066998903198   -0.0000002309632277   12.4200976915162187

  C   N   H   Pb   I

   4   8   20   4   12

Direct

 0.7746435153409925  0.2746449060650629  0.2003670471117680

 0.2711261399878776  0.7711248044750924  0.3007342335893508

 0.7746430580767620  0.2746450323147297  0.7996317557940844

 0.2711257386760722  0.7711223563130301  0.6992659855842422

 0.6816357376202636  0.1816363191025259  0.2475647454306364

 0.1782194550161230  0.8641466013626504  0.2534499672510961

 0.6816349665263020  0.1816364299470777  0.7524342007467655

 0.1782188863469763  0.8641443683557698  0.7465503757549949

 0.8676578618391081  0.3676584617663631  0.2474699477030123

 0.3641477228628140  0.6782180737527591  0.2534498331818164

 0.8676575687691330  0.3676587753664792  0.7525292693297934

 0.3641475178650473  0.6782159389516810  0.7465502466613771

 0.8799445494498562  0.3799443704599036  0.3297223727991794

 0.3766992796960714  0.6663797150286365  0.1711603679175100

 0.8799443579873951  0.3799442013574919  0.6702768981844190

 0.3766999121297549  0.6663778588717880  0.8288396813718298

 0.9344810754575056  0.4344818298124393  0.1990459618909138

 0.4310582020585884  0.6112024150975233  0.3016504197603562

 0.9344810589706463  0.4344816199178362  0.8009536728790765

 0.4310591191893166  0.6112011833655606  0.6983496601732597

 0.6693680407373164  0.1693678292004197  0.3298118958450473

| | | |
|---|---|---|
| 0.1663806942996040 | 0.8766984655659876 | 0.1711604258409424 |
| 0.6693673531279465 | 0.1693677931793632 | 0.6701871617109356 |
| 0.1663811628229934 | 0.8766965196133091 | 0.8288397780088764 |
| 0.6147228373430703 | 0.1147236390125636 | 0.1992684927730775 |
| 0.1112037981567041 | 0.9310574012509422 | 0.3016504280881107 |
| 0.6147225772068775 | 0.1147232435184856 | 0.8007304677863140 |
| 0.1112044773297577 | 0.9310559643523985 | 0.6983497241292345 |
| 0.7745300924432077 | 0.2745320428547652 | 0.1115531261797778 |
| 0.2707173401039470 | 0.7707161873408765 | 0.3895207253702290 |
| 0.7745299833791898 | 0.2745319712492315 | 0.8884459579814887 |
| 0.2707173139570478 | 0.7707142549114633 | 0.6104791846648372 |
| 0.2727911064503869 | 0.2716471586547899 | 0.4999999582970716 |
| 0.2732369289931407 | 0.2722687948180227 | 0.9999999482394357 |
| 0.7716477056416657 | 0.7727906527675341 | 0.5000004497242211 |
| 0.7722691155964032 | 0.7732367873267253 | 0.0000004294655155 |
| 0.5470691086499139 | 0.0470677489878385 | 0.4999995337610246 |
| 0.0020175083391995 | 0.5020159322672415 | 0.5000001647324872 |
| 0.5070719526450486 | 0.5385138383964089 | 0.4999999409777253 |
| 0.0385168234235788 | 0.0070689153527740 | 0.5000003161629971 |
| 0.5428653751600229 | 0.0428648495033139 | 0.9999998385324955 |
| 0.0072381229315927 | 0.5072374909853751 | 0.9999996203292063 |
| 0.4987143269590430 | 0.5479050038254458 | 0.0000004333818766 |
| 0.0479069694746347 | 0.9987124778232942 | 0.9999997989052872 |
| 0.2735032906527326 | 0.2713816740976678 | 0.7500031798824776 |
| 0.2735035310747365 | 0.2713833127161021 | 0.2499965630375837 |
| 0.7713698963810640 | 0.7735073398892371 | 0.7500045865788579 |
| 0.7713828968526043 | 0.7735033288539945 | 0.2499981864973736 |

CsPbI₃-orthorhombic

  1.02000000000000

   8.9599414455581332   0.0000000000000000  -0.0019858272315529

   0.0000000000000000  12.2225883518336147   0.0000000000000000

  -0.0017217251913990   0.0000000000000000   7.9327952756357982

  Cs  Pb  I

   4   4   12

Direct

 0.0969151594643876 0.2500000000000000 0.9530078951339789

 0.9033814508768998 0.7500000000000000 0.0467686762672770

 0.4033051129230714 0.7500000000000000 0.4532687677607186

 0.5970795436661973 0.2500000000000000 0.5468552783607521

 0.5000457781896586 0.4999462016998493 0.0000371185473611

 0.0001531021329768 0.4999511693573453 0.5001021746821408

 0.5000457781896586 0.0000537983001507 0.0000371185473611

 0.0001531021329768 0.0000488306426547 0.5001021746821408

 0.1863979457335461 0.5399207631884479 0.1694683722032480

 0.8137584223543968 0.4600197733542970 0.8304861121995089

 0.3137272263034205 0.4602277904076715 0.6694774311414946

 0.6865060990948422 0.5396946570264802 0.3304969521010790

 0.8137584223543968 0.0399802266457030 0.8304861121995089

 0.1863979457335461 0.9600792368115521 0.1694683722032480

 0.6865060990948422 0.9603053429735198 0.3304969521010790

 0.3137272263034205 0.0397722095923285 0.6694774311414946

 0.4960645257050089 0.2500000000000000 0.0835340882284186

 0.5040218884636261 0.7500000000000000 0.9168994709025000

 0.0036421718080888 0.7500000000000000 0.5829772433238887

 0.9963673345995687 0.2500000000000000 0.4165522532728048

----------------------------------------------------------------------------------------------------

CsPbI3-cubic

   1.02500000000000

   6.1349999999999998     0.0000000000000000     0.0000000000000000

   0.0000000000000000     6.1349999999999998     0.0000000000000000

   0.0000000000000000     0.0000000000000000     6.1349999999999998

  Cs   Pb   I

   1    1    3

Direct

  0.0000000000000000  0.0000000000000000  0.0000000000000000

  0.5000000000000000  0.5000000000000000  0.5000000000000000

  0.0000000000000000  0.5000000000000000  0.5000000000000000

  0.5000000000000000  0.0000000000000000  0.5000000000000000

  0.5000000000000000  0.5000000000000000  0.0000000000000000

----------------------------------------------------------------------------------------------------

FASnI3-cubic

  1.035000000000000

     6.2201695923162061   -0.0000000282245871    0.0000000000000000

     0.0000007823812975    6.0530228143591973    0.0000000000000000

     0.0000000000000000    0.0000000000000000    6.1334433839192787

  C   N   H   Pb   I

   1    2    5    1    3

Direct

  0.5000010130040877  0.5286908332526608  0.5000000000000000

  0.6860491372913060  0.4306492098538683  0.5000000000000000

  0.3139488957057068  0.4306492098453205  0.5000000000000000

  0.5000000000030340  0.7104787945879707  0.5000000000000000

  0.8202393923822626  0.5293663220288636  0.5000000000000000

  0.7071705157137629  0.2620558265487887  0.5000000000000000

  0.2928294842809933  0.2620558265453923  0.5000000000000000

  0.1797606076200182  0.5293663220208842  0.5000000000000000

  0.9999999999987921  0.9497696661704040  0.0000000000000000

  0.4999999999983515  0.9640622716905227  0.0000000000000000

  0.9999999999986855  0.4466878132019190  0.0000000000000000

  0.0000000000029985  0.9246828552534225  0.5000000000000000

---------------------------------------------------------------------------------------------------------

FASnI₃-orthorhombic

```
  1.035000000000000
   8.5658459056244389    0.0338022583601564   -0.0385878766465057
   0.0559293510024624   12.0968766655331041    0.0253851807896190
  -0.0394434188128824    0.0196393360834740    8.7158064637311110
  Pb  I   C   N   H
   4  12   4   8  20
Direct
 0.5079511562182449 0.0005797799328977 0.4963668928576073
 0.0155538576678002 0.0007092725349378 0.0034547503551771
 0.5066096732666163 0.4998538986955256 0.4973032185935099
 0.0128526658943417 0.5013733327859744 0.0018227789997680
 0.5113542293145144 0.2512851169795850 0.4904284002785421
 0.5059690770703408 0.7511311892462434 0.4881087259278313
 0.0243455039350082 0.7510740119262402 0.0061380114365411
 0.0288120170319063 0.2509308812014363 0.9969089462369313
 0.7814361603747821 0.0009381950567905 0.7309769922389656
 0.2176924341209736 0.9992162014942775 0.2918684139757194
 0.7292540268309380 0.0112998975694466 0.2185098544134861
 0.2827490930947780 0.0036106115106566 0.7703808896903863
 0.2140238214000908 0.4959377594706731 0.2936834769705606
 0.7825287364763812 0.5004218689801739 0.7278481613978846
 0.2842333798993784 0.5075994784683050 0.7735941600065852
 0.7244796822905446 0.5086980518810619 0.2159666712553618
 0.0015618845088048 0.7095239898679726 0.5020925333506431
 0.5151632290557520 0.7279254156190906 0.9910048186091903
 0.0034003616205402 0.2090368011988994 0.4963023605736533
 0.5110943423121466 0.2269660125447352 0.9874024364038462
 0.9180429339470949 0.7612911809510345 0.3982550797521072
```

| | | |
|---|---|---|
| 0.1014766470065179 | 0.7545004070009312 | 0.5960621512416249 |
| 0.6184437288726743 | 0.7518658172496643 | 0.8876715547977719 |
| 0.4441055051558694 | 0.7997762972713105 | 0.0779427020794147 |
| 0.9231522499750777 | 0.2584049529224399 | 0.3878664874321139 |
| 0.1055980768338299 | 0.2556991229372042 | 0.5862111575958121 |
| 0.6238677204569060 | 0.2516974576981043 | 0.8944854351956344 |
| 0.4401208222164990 | 0.2967423982828125 | 0.0774414902538855 |
| 0.9851863789898407 | 0.6198598206042034 | 0.5128136155793428 |
| 0.8427092295539598 | 0.7165173131488018 | 0.3310262999001896 |
| 0.9246670133243964 | 0.8448902593007660 | 0.3776103784997663 |
| 0.1269084711267568 | 0.8375111146452249 | 0.5974729927821338 |
| 0.1620140875601551 | 0.7039923535819164 | 0.6716546674729530 |
| 0.4851458232197488 | 0.6402973896850286 | 0.0054707987415316 |
| 0.6694590125852834 | 0.6885852475646916 | 0.8261415889057282 |
| 0.6530220232218849 | 0.8316784274403908 | 0.8628456296706062 |
| 0.4683274748515552 | 0.8827857189656038 | 0.0752037909033518 |
| 0.3636837498560751 | 0.7743744333836242 | 0.1565399287828380 |
| 0.9814801885500501 | 0.1206711647951993 | 0.5149620311442413 |
| 0.8453078342819831 | 0.2128003134851436 | 0.3242985061239703 |
| 0.9359009605368865 | 0.3404750776081850 | 0.3597425721462459 |
| 0.1354072458227803 | 0.3377240542971691 | 0.5785033322894793 |
| 0.1629498864461515 | 0.2078779404470268 | 0.6671127973568727 |
| 0.4719283761854801 | 0.1406104876417028 | 0.9886767201922748 |
| 0.6747062591178413 | 0.1896663812866801 | 0.8303719842532724 |
| 0.6641251691549479 | 0.3312262629103195 | 0.8784860481495307 |
| 0.4731789298150988 | 0.3779523015035389 | 0.0873058331243811 |
| 0.3476386405692209 | 0.2716199404818598 | 0.1421616156357852 |

----------------------------------------------------------------------------------------------------

CsSnI$_3$-orthorhombic

  1.03000000000000

   8.6487088099199845    0.0000000000000000   -0.0005590576585399

   0.0000000000000000   12.0732946125359923    0.0000000000000000

  -0.0004853281656927    0.0000000000000000    8.1898216167006659

  Cs   I   Sn

   4   12   4

Direct

 0.0728370918164529 0.2500000000000000 0.9751148755566916

 0.9274120969952477 0.7500000000000000 0.0248992454268873

 0.4275560972050485 0.7500000000000000 0.4750731272257482

 0.5728082979335483 0.2500000000000000 0.5249022965552257

 0.1982407028883131 0.5329258059963067 0.1925440385115635

 0.8019043034626421 0.4670859241900800 0.8075523916553990

 0.3019164866063093 0.4670798223126624 0.6923117597735668

 0.6982497416491213 0.5327676566524886 0.3075335707549627

 0.8019043034626421 0.0329140758099200 0.8075523916553990

 0.1982407028883131 0.9670741940036933 0.1925440385115635

 0.6982497416491213 0.9672323433475114 0.3075335707549627

 0.3019164866063093 0.0329201776873376 0.6923117597735668

 0.4985614870107824 0.2500000000000000 0.0643105106349040

 0.5014495294247894 0.7500000000000000 0.9359283205763234

 0.0014746047071625 0.7500000000000000 0.5644281321087945

 0.9990966082302606 0.2500000000000000 0.4354144395121295

 0.5000141384432126 0.4999955501547930 0.0000043047820810

 0.0000538878510312 0.4999900253457383 0.5000184582240692

 0.5000141384432126 0.0000044498452070 0.0000043047820810

 0.0000538878510312 0.0000099746542617 0.5000184582240692

CsSnI3-cubic
```
  1.00000000000000
     6.2190000000000003    0.0000000000000000    0.0000000000000000
     0.0000000000000000    6.2190000000000003    0.0000000000000000
     0.0000000000000000    0.0000000000000000    6.2190000000000003
   Cs   Pb   I
    1    1    3
Direct
  0.0000000000000000  0.0000000000000000  0.0000000000000000
  0.5000000000000000  0.5000000000000000  0.5000000000000000
  0.0000000000000000  0.5000000000000000  0.5000000000000000
  0.5000000000000000  0.0000000000000000  0.5000000000000000
  0.5000000000000000  0.5000000000000000  0.0000000000000000
```

---------------------------------------------------------------------------------------------------------------

MASnI3-cubic

  1.02500000000000

    6.0731745396124639   -0.0010743309947183    0.0300803266638338

   -0.0010786467260402    6.0629924287505892    0.0002089676266765

    0.0304395621010224    0.0002498487131757    6.1330237940473760

  C   N   H   Sn   I

  1   1   6   1   3

Direct

 0.8987365275931865  0.0000134481207965  0.9909359249804979

 0.1358550153965581  0.9997785402548942  0.0354665914758385

 0.8726758937441303  0.9997515053307495  0.8134723120554028

 0.8240698409334186  0.1488681039933297  0.0635942238500746

 0.8237021675801657  0.8513901295750443  0.0640011724487266

 0.2159546615323151  0.1391089090523536  0.9714449345518190

 0.2156096903431433  0.8602234220085876  0.9717698142589697

 0.1705972267925233  0.9999862472942453  0.2026158709095540

 0.4806462343774882  0.4999609390329525  0.4786849076973851

 0.4485867841010034  0.4999303196444984  0.9731206214471015

 0.4491222611188945  0.0000280810340598  0.5179706166544307

 0.9662527254871733  0.5002313646584824  0.4370639996702081

MASnI₃-tetragonal

  1.025000000000000

   8.4324703751439785   -0.0014233652348497   -0.0005462219600508

   0.0051493342194156    8.4418034305723282    0.0007980490596414

   0.1085370329072245    0.0822098583503458   12.3771642636398145

  C   N   H   Sn   I

  4   4   24   4   12

Direct

 0.0310047763814083  0.4967442625037322  0.2505600656488980

 0.5333594200491021  0.0470699605841460  0.2502322783296691

 0.0465120456555894  0.5337812256295180  0.7496201763884827

 0.4970171865202317  0.0314093780598981  0.7510493721858325

 0.9259826960118076  0.6085452125758906  0.1949242829322557

 0.4278746098154897  0.9334020434775425  0.1979940161317515

 0.9341235768647635  0.4267925345756396  0.6975702219705724

 0.6080281578640694  0.9253309511273500  0.6952965355739025

 0.6572219593299238  0.0209264807512355  0.2273345459305958

 0.5187171583604027  0.0369295383936787  0.3384362002544776

 0.5026065307618239  0.1677889312971885  0.2232062062813895

 0.4529659050294796  0.8159998891977551  0.2206382430241760

 0.4391685271748642  0.9396288706228830  0.1139882105489249

 0.3081094647283606  0.9552801932374493  0.2167454795156658

 0.9355926759816242  0.5970864788257586  0.1111964361251623

 0.9526264120490566  0.7268607189295153  0.2133088727503463

 0.8062665943215208  0.5888632654168404  0.2153010095069519

 0.1547851126535846  0.5201680150670995  0.2260995024251287

 0.9987370782305973  0.3748676831963138  0.2284147444862867

 0.0174669682785407  0.5117640763855462  0.3383510751689158

 0.9544189015692197  0.3078876451827313  0.7189291747538462

| | | |
|---|---|---|
| 0.9439743924307322 | 0.4339051023583451 | 0.6135206255086203 |
| 0.8158725494491748 | 0.4534969774980979 | 0.7177626560817032 |
| 0.1681041397934777 | 0.5011416063913856 | 0.7252690517306633 |
| 0.0329563159628350 | 0.5230973816268474 | 0.8378278313120333 |
| 0.0220646687802315 | 0.6564633721156952 | 0.7238279627486364 |
| 0.7267142894597143 | 0.9526437561029582 | 0.7122535204868541 |
| 0.5946819683705797 | 0.9327408944958648 | 0.6116461279232652 |
| 0.5888665761440350 | 0.8061890629306632 | 0.7170669900323148 |
| 0.5198342882072495 | 0.1544687455895826 | 0.7249789191016163 |
| 0.5136141066201816 | 0.0200895235859733 | 0.8388054519073194 |
| 0.3745850558561656 | 0.9985988239305996 | 0.7304078268470278 |
| 0.5089527856428049 | 0.5163653881797003 | 0.9891270880605916 |
| 0.5166528083061124 | 0.5110793488880603 | 0.4890399879855565 |
| 0.0283329705763293 | 0.0118523621561977 | 0.4909601333116811 |
| 0.0104846587221274 | 0.0288759166824946 | 0.9908398646389998 |
| 0.7047481228002681 | 0.8304468606653046 | 0.9925693187097977 |
| 0.3225450108146006 | 0.2126770639553399 | 0.9958997780225047 |
| 0.2098542513729740 | 0.7203087350874497 | 0.9968145339335379 |
| 0.8284423656147837 | 0.3395700749544019 | 0.9888648828066309 |
| 0.3396551477951704 | 0.8286639232002955 | 0.4890767479195119 |
| 0.7199853397241469 | 0.2102192291682172 | 0.4972326941013634 |
| 0.8312969442323848 | 0.7053467417230550 | 0.4928168833521411 |
| 0.2130553677478844 | 0.3229312741213377 | 0.4949241378956444 |
| 0.5184183268427276 | 0.5229839970113588 | 0.2407642345638976 |
| 0.5231117898137398 | 0.5181837569921868 | 0.7407824583424159 |
| 0.0159022310533246 | 0.0165548305338490 | 0.7428864818175640 |
| 0.0167060452347911 | 0.0159738810170182 | 0.2428350099248178 |

MASnCl3-cubic

   1.02500000000000

   5.4963258491905451    0.0008873735280529    0.1469738718269263

   0.0008399546602409    5.4018715988923960    0.0006809677306016

   0.1497648062003817    0.0007048447993364    5.5638535750055951

  C   N   H   Sn   Cl

   1   1   6   1   3

Direct

 0.8890755384075462  0.9997217919795034  0.9684461698724363

 0.1361112083035323  0.9997067393830221  0.0549990260238076

 0.8993397909070140  0.9995136347822040  0.7703556972025112

 0.7921043273519146  0.1669917935783118  0.0338680010240253

 0.7920639972485617  0.8326403248275867  0.0342853591746106

 0.2342580670585477  0.1569490962901483  0.9953832115191261

 0.2342690949236399  0.8423935658903758  0.9953765974498694

 0.1400921478243617  0.9995611872015360  0.2423348390470466

 0.4808368436361334  0.5004199081197243  0.4761663997061092

 0.4364930071059732  0.4996559770751290  0.9513545236941354

 0.4383244536704325  0.0002831939151875  0.5240661604268055

 0.0050129115622966  0.5014652339572194  0.4491517548595496

-------------------------------------------------------------------------------------------------------

MASnCl3-orthorhombic

  1.02500000000000

   7.2362927052732058   0.0000000000000000   0.0000000000000000

   0.0000000000000000  10.9610728622243681   0.0000000000000000

   0.0000000000000000   0.0000000000000000   8.0981980944705665

  C   N   H   Sn   Cl

   4   4   24   4   12

Direct

  0.5352495017594450  0.2500000000000000  0.9224402873250739

  0.4647504982405550  0.7500000000000000  0.0775597126749190

  0.9647504982405550  0.7500000000000000  0.4224402873250810

  0.0352495017594450  0.2500000000000000  0.5775597126749261

  0.4639420312109763  0.2500000000000000  0.0921300215030598

  0.5360579687890237  0.7500000000000000  0.9078699484969448

  0.0360579687890237  0.7500000000000000  0.5921300515030552

  0.9639420912109742  0.2500000000000000  0.4078699784969402

  0.5111392443440366  0.3271600075785486  0.1558891198375321

  0.4888607556559634  0.6728400224214539  0.8441108651624702

  0.9888607556559634  0.6728400224214539  0.6558891348375298

  0.0111393033440379  0.3271600075785486  0.3441108651624702

  0.4888607556559634  0.8271599775785461  0.8441108651624702

  0.5111392443440366  0.1728400074214562  0.1558891198375321

  0.0111393033440379  0.1728400074214562  0.3441108651624702

  0.9888607556559634  0.8271599775785461  0.6558891348375298

  0.3191378764291741  0.2500000000000000  0.0916569900283690

  0.6808621235708259  0.7500000000000000  0.9083430399716264

  0.1808621235708259  0.7500000000000000  0.5916569600283736

  0.8191378764291741  0.2500000000000000  0.4083430099716310

  0.4852372898530746  0.3322194169958763  0.8582720630511957

0.5147627101469254  0.6677806130041191  0.1417279369488043

0.0147627101469254  0.6677806130041191  0.3582720630511957

0.9852372898530746  0.3322194169958763  0.6417289499488064

0.5147627101469254  0.8322193869958809  0.1417279369488043

0.4852372898530746  0.1677805980041214  0.8582720630511957

0.9852372898530746  0.1677805980041214  0.6417289499488064

0.0147627101469254  0.8322193869958809  0.3582720630511957

0.6875187166861707  0.2500000000000000  0.9260305204597401

0.3124812833138293  0.7500000000000000  0.0739694575402581

0.8124812833138293  0.7500000000000000  0.4260305504597426

0.1875187166861707  0.2500000000000000  0.5739694795402599

0.5000000000000000  0.0000000000000000  0.5000000000000000

0.0000000000000000  0.0000000000000000  0.0000000000000000

0.5000000000000000  0.5000000000000000  0.5000000000000000

0.0000000000000000  0.5000000000000000  0.0000000000000000

0.5381598210221767  0.2500000000000000  0.5078098013613186

0.4618401789778233  0.7500000000000000  0.4921901686386789

0.9618401789778233  0.7500000000000000  0.0078098303613174

0.0381598210221767  0.2500000000000000  0.9921901986386814

0.7927599039183377  0.9737351626425692  0.7147127078373288

0.2072400960816623  0.0262648313574303  0.2852872921626712

0.7072400960816623  0.0262648313574303  0.2147127228373336

0.2927599039183377  0.9737351626425692  0.7852872921626712

0.2072400960816623  0.4737351626425692  0.2852872921626712

0.7927599039183377  0.5262648373574308  0.7147127078373288

0.2927599039183377  0.5262648373574308  0.7852872921626712

0.7072400960816623  0.4737351626425692  0.2147127228373336